\documentstyle[preprint,revtex]{aps}
\begin{document}
\draft
\preprint{Appeared in PRB {\bf 46},4663 (1992)}
\preprint{September 1992}
\begin{title}
Spectrum and Thermodynamics
of the 1D
Supersymmetric t-J\\
Model with $1/r^2$ Exchange and Hopping
\end{title}
\author{D. F. Wang}
\begin{instit}
Joseph Henry Laboratories of Physics, Princeton University, \\
Princeton, New Jersey 08544
\end{instit}
\author{James T. Liu}
\begin{instit}
Institute of Field Physics, University of North Carolina, \\
Chapel Hill, North Carolina 27599-3255
\end{instit}
\author{P. Coleman}
\begin{instit}
Serin Laboratories, Rutgers University, \\
P.O.~Box 849, Piscataway, New Jersey 08854
\end{instit}
\begin{abstract}
We derive the spectrum and thermodynamics
of the 1D supersymmetric t-J model with long
range hopping and spin exchange using a set of maximal spin
eigenstates.  This spectrum  confirms the recent conjecture that the
asymptotic Bethe-ansatz spectrum is exact. By empirically
determining the spinon degeneracies of each state, we are
able to explicitly construct the free energy.
\end{abstract}

\narrowtext

\def\hh{{\sl \Phi_{\vphantom{h}s}}}
\def\ss{{\sl \Phi_h}}
The explicit construction of low dimensional
models with Jastrow ground-state wavefunctions has attracted
considerable recent interest\cite{laughlin,girvin,arovas,mele,hald88,shas88}.
In one dimension, Shastry and Haldane\cite{hald88,shas88} have demonstrated
that the ground-state of the 1D Heisenberg model with a $1/r^2$
exchange interaction is a Gutzwiller state for the half filled
infinite-$U$ Hubbard model.  Haldane has shown how the spectrum of
this model can be written in terms of a generalized type of Jastrow
wavefunction with excitations of novel statistics\cite{hald91}.

Kuramoto and Yokoyama\cite{kura91} have recently extended these
results to include holes, demonstrating that the corresponding 1D
supersymmetric t-J model is also characterized by a Gutzwiller
ground-state.  Most recently, Kawakami has obtained an asymptotic
Bethe-ansatz (ABA) solution for the model, based on the observation
that the ground-state wavefunction is a  product of two-body
functions\cite{kawa91}.  Assuming factorizability, he derived the
spectrum of the system, which was conjectured to be exact.  The
low-energy critical behavior of the model has been identified as a
Luttinger liquid\cite{yang91,Kawakami92}; the spin and charge excitations are
described independently by $c=1$ conformal field theories.

In the case of the $1/r^2$ Bose gas\cite{suth72}, and the Shastry-Haldane
$1/r^2$
Heisenberg chain\cite{hald91}, the ABA has been shown to furnish the correct
spectrum,
despite the long-range nature of the interactions.
A remarkable feature of these models is that excited states
are obtained from the ground-state
by introducing zeros into the Jastrow wavefunction,
in a manner reminiscent of Laughlin's description of
quasiparticles in the fractional quantum Hall effect.
This motivates us to examine
the $1/r^2$ supersymmetric t-J model in a
similar vein.  Here, we show how this
philosophy can be used to
construct the excited state Jastrow wavefunctions of the $1/r^2$
supersymmetric t-J model and indeed, the spectrum
confirms Kawakami's  conjecture.
In addition to the spectrum, we are able to obtain the spin
degeneracies of each state, permitting us to write the
the free energy in closed form.


The Hamiltonian for the one-dimensional t-J model is given by
\begin{equation}
H=\sum_{{i\ne j}, \sigma}\left[ -t_{ij} c_{i\sigma}^\dagger
c_{j\sigma}\right] +\sum_{i\ne j}\left[ J_{ij}({\bf S}_i\cdot{\bf S}_j
 -{\textstyle{1\over4}}n_i n_j) \right],
\label{eq:hamil}
\end{equation}
%
%
where we implicitly project out any double occupancies.
We take $t_{ij}=J_{ij}=t/d^2(i-j)$ where $d(n)={N\over\pi}\sin(n\pi/N)$
is the chord distance consistent with periodic boundary conditions on
$N$ lattice sites\cite{Rucken91}.
\def\up{\uparrow}
\def\down{\downarrow}

States in the Hilbert space can be represented by spin and hole
excitations from the fully-polarized up-spin state
$|P\rangle$\cite{anderson89}.  If
we let $Q$ denote the number of holes and $M$ denote the number of
down-spins, then $S_z$ is given by $S_z=(N-Q)/2-M$.  The wavefunctions
are given by
\begin{equation}
|\psi\rangle=\sum_{x,y}\psi(x,y)
\prod_\alpha S^{-}_{x_\alpha} \prod_i h_{y_i}^\dagger
|P\rangle,
\label{eq:wave}
\end{equation}
where the amplitude $\psi(x,y)$ is symmetric in
$x \equiv (x_1,x_2,\ldots,x_M)$, the positions of the down-spins,
and antisymmetric in $y \equiv ( y_1,y_2,\ldots,y_Q)$,
the positions of the holes.
$S^{-}_{x_{\alpha}}=c_{x_\alpha\down}^\dagger
c_{x_\alpha\up}^{\vphantom{\dagger}}$ is the spin-lowering operator
at site $x_\alpha$ and $h_{y_i}^\dagger=c_{y_{i\up}}$ creates a hole at
site $y_i$.

We can construct a general class of states corresponding to states of
uniform motion and spin polarization.  To describe these states, we
generalize Kuramoto and Yokoyama's Jastrow ground-state\cite{kura91}
as follows
\widetext
\begin{eqnarray}
\psi_G(x,y;J_s,J_h)=&&\exp\left[{2\pi i\over N}
\left({J_s\sum_\alpha x_\alpha}+{J_h\sum_i y_i}\right)\right]{\bf
 \Psi}_0(x,y)\nonumber\\
\quad {\bf \Psi}_0(x,y)=&&
\prod_{\alpha<\beta}d^2(x_\alpha-x_\beta)\prod_{i<j}d(y_i-y_j)
\prod_{\alpha,i}d(x_\alpha-y_i).
\label{eq:gutz}
\end{eqnarray}
\narrowtext
Here,  $J_s$ and $J_h$ govern the (uniform) momenta of
down-spins and holes respectively.  $J_s$ and $J_h$ take on either
integral or half-integral values as appropriate to insure that
$\psi_G$ has the correct periodicities under $x_\alpha \to x_\alpha+N$
and $y_i \to y_i+N$.

The Hamiltonian can be broken up into four parts,
$H=T^\up+T^\down+H^0+H^{\rm int}$,
where $T^\up$ ($T^\down$) is the up (down) spin transfer operator,
$H^0$ is the spin exchange operator and $H^{\rm int}$ is the diagonal
interaction term.  When $H$ acts on $\psi_G$, $T^\up$ only affects the
$y$ variables and $H^0$ only affects the $x$ variables.  As a
result, these operators are easy to treat and yield only two and three
body terms when appropriate conditions on $J_s$ and $J_h$ are
met\cite{hald88,shas88,kura91}.

However, because $T^\down$ exchanges
pairs of $x_\alpha$ and $y_i$, this term must be treated differently.
In general, it is {\it not} true that $T^\up|\psi_G\rangle
=T^\down|\psi_G\rangle$ because $T^{\up}$ does not commute with
the spin raising operator. This difficulty was overlooked in earlier
work\cite{kura91}.
To deal with $T^\down$, we use an alternate representation for
$|\psi_G\rangle$ in terms of up-spins and holes.  Let us
introduce the $N-M-Q$ coordinates $u \equiv
(u_1,u_2,\ldots,u_{N-M-Q})$ which give the location of
the up-spins.  Wavefunctions in this representation are given by the
spin rotated version of Eq.~(\ref{eq:wave}) where the $x$ are
replaced by $u$ and $M$ is replaced by $N-M-Q$.  Making this transformation,
we find
\begin{equation}
\psi_G(x,y;J_s,J_h)=A\psi_G(u,y;N-J_s,J_h-J_s+{N\over2}),
\label{eq:dual}
\end{equation}
where the set of $N$ coordinates $(x,y,u)$ exhausts the entire lattice.
$A$ is a
constant independent of the spin and hole coordinates.
Using this identity, the down-spin transfer operator gives
\begin{equation}
{T^\down\psi_G(x,y)\over\psi_G(x,y)}=
{T^\down\psi_G(u,y)\over\psi_G(u,y)},
\end{equation}
and can thus be treated in a similar manner as $T^\up$.  The result
gives two and three body terms in the variables $u$ and $y$.
These terms can then be converted into sums over the $x$ and $y$
variables by making use of the fact that $(x,y,u)$ runs over the
entire lattice.

When the separate terms that contribute to the Hamiltonian are
combined, we find that the two body terms drop out and the three body
terms combine to give constants.  As a result, $\psi_G$ with total momentum
$P={2\pi\over N}(J_s M + J_h Q)$ is an exact eigenstate of $H$ with energy
\begin{eqnarray}
{N^2\over\pi^2t}E=&&{2\over3}M(M^2-1)-2MJ_s(N-J_s)\nonumber\\
&&+Q\biggl[{1\over3}(N^2-1)+{2\over3}(Q^2-1)+{1\over2}(M+Q)(2M-Q)\nonumber\\
&&\quad\quad-2J_h(N-J_h)+2(J_s-J_h)^2\biggr].
\label{eq:energy}
\end{eqnarray}
The cancellation of the many body terms, and thus this result, is only
valid under the conditions
$|J_s-N/2|\le N/2-(M-1+Q/2)$,
$|J_h-N/2|\le N/2-(M+Q-1)/2$ and
$|J_h-J_s|\le(M+1)/2$.
For a given $S_z$, the minimum energy is given when $J_s$ and $J_h$
are as close to $N/2$ as possible.  The ground state is given when
$S_z$ is either $0$ or $1/2$ and is a singlet whenever
possible\cite{kura91}.  When $Q=0$ this reduces to the result for the
Heisenberg chain\cite{hald88,shas88}.
These energy levels have also been found by Ha and Haldane\cite{ha92},
where $J_{\uparrow} = J_h -N+(M+Q+1)/2$ and $J_{\downarrow} =
J_h - J_s - (M-1)/2$. From these energy levels we find the spin
and charge velocities to be identical to the previous
results\cite{kura91,ha92}.


To investigate the other excited states of the system,
we introduce zeros into the  wavefunction by premultiplying
it with polynomials of $X_\alpha=\exp (2\pi i x_{\alpha} /N)$ and
$Y_i=\exp (2\pi i y_i /N)$.
The wavefunctions thus take the following modified
Kalmeyer-Laughlin
form\cite{kalm87}:
\widetext
\begin{equation}
\psi(x,y)=\hh (X,Y)\ss (Y) {\bf \Psi_0},\end{equation}
\narrowtext
where $\hh $ and $\ss$ are  completely symmetric under pairwise interchange
of their arguments.  These states will be termed ``fully-polarized spinon
states''.  Loosely speaking, the polynomials $\hh $ and $\ss $ can be
regarded as spin and charge quasiparticle wavefunctions respectively.

When the Hamiltonian acts on this wavefunction, once again all
three-body terms combine to give constants.  However, in this case,
some two-body terms remain and we are left with the eigenvalue
equation
\begin{equation}
{N^2\over\pi^2t}E\hh \ss  = E_0\hh \ss  + H_1 + H_2 + H_3,
\label{eq:eval}
\end{equation}
where
\begin{eqnarray}
H_1 &=&2\ss \left[\sum_\mu\partial_\mu^2+\sum_{\mu<\nu} {W_\mu+W_\nu
\over W_\mu-W_\nu}(\partial_\mu-\partial_\nu)\right]\hh \nonumber\\
&&+4\hh \left[\sum_i\partial_i^2+{1\over2}\sum_{i<j} {Y_i+Y_j \over
Y_i-Y_j}(\partial_i-\partial_j)\right]\ss \nonumber\\
H_2 &=&4\sum_i\partial_i\hh \partial_i\ss \nonumber\\
H_3 &=&2\ss \sum_{\alpha<\beta}{X_\alpha+X_\beta\over
X_\alpha-X_\beta}(\partial_\alpha-\partial_\beta)\hh .
\end{eqnarray}
Here  we denote $W\equiv(X,Y)\equiv(X_1 \dots X_M, Y_1 \dots Y_{Q+M})$ and
$\partial_\mu\equiv
W_\mu\partial/\partial W_\mu$.  In deriving this, we have shifted $\hh $
by the ground state configuration, $\prod_\mu W_\mu^{N/2}$.  As a result,
$E_0$ is given by Eq.~(\ref{eq:energy}) with $J_s=J_h=N/2$.
We require that
$|{\rm degree}\>\hh |\le N/2-(M-1+Q/2)$,
$|{\rm degree}\>\hh \ss |\le N/2-(M+Q-1)/2$ and
$|{\rm degree}\>\ss |\le(M+1)/2$,
which is to hold for each variable $X_\alpha$ or $Y_i$ independently.

\def\nn{\{n\}}
\def\mm{\{m\}}
The first term, $H_1$, does not mix $\hh $ and $\ss $ and has been
solved by Sutherland\cite{suth72}.  However, $H_2$ mixes $\hh $ and
$\ss $ and $H_3$ does not act symmetrically on $\hh $.  As a
result, they are harder to deal with.  We follow Sutherland and start
by choosing the following symmetric basis functions:
\begin{eqnarray}
\hh(W;\nn)&=&\sum_{\{P_\mu\}}
\prod_\mu W_\mu^{n_{P_\mu}}\nonumber\\
\ss(Y;\mm)&=&\sum_{\{{P_i}\}}
\prod_i Y_i^{m_{{P_i}}},
\label{eq:basis}
\end{eqnarray}
where the quantum numbers $\{n_1,\ldots,n_{M+Q}\}$ and
$\{m_1,\ldots,m_Q\}$ are taken to be in increasing order and
$\{P_{\mu}\}$ and $\{ P_i\}$ denote permutations of the indices.
These quantum numbers are integral or
half-integral as required by periodic boundary conditions.

\def\sgn{{\rm sgn}}
In this basis, labeled by the two sets of quantum numbers $\{n_\mu\}$
and $\{m_i\}$, the Hamiltonian, considered as a matrix,
can be shown to be upper-triangular.  Eigenvalues are
found by reading the diagonal-elements labeled in terms of the
quantum numbers $\{n_\mu\}$ and $\{m_i\}$\cite{unpub}.
The result simplifies when written
in terms of a conjugate set of quantum numbers
$\{J_1,J_2,\ldots,J_{M+Q}\}$ and $\{I_1,I_2,\ldots,I_Q\}$ defined by
\begin{eqnarray}
J_\mu&=&n_\mu+n_\mu^0\qquad\quad\ \, n_\mu^0={1\over2}(2\mu-(M+Q)-1)\nonumber\\
I_i&=&m_i+m_i^0\qquad\quad   m_i^0={1\over2}(2i-Q-1),
\end{eqnarray}
where $\{n_\mu\}$ and $\{m_i\}$ must satisfy the conditions
specified before.  This translates into the conditions $|J_\mu|\le
(N-M+1)/2$ and $|I_i|\le (M+Q)/2$.
The energy is
\begin{equation}
{E\over t}={\pi^2\over3}Q(1-{1\over N^2})+{1\over2}\sum_{\mu=1}^{M+Q}
(p_\mu^2-\pi^2),
\label{eq:eigen}
\end{equation}
where the pseudomomenta, $p_\mu$, are given by the following equations:
\begin{eqnarray}
p_\mu N=2\pi J_\mu-\pi&&\sum_{i=1}^Q \sgn(p_\mu-q_i)
+\pi\sum_{\nu=1}^{M+Q}\sgn(p_\mu-p_\nu)\nonumber\\
2\pi I_i=\pi&&\sum_{\mu=1}^{M+Q}\sgn(q_i-p_\mu).
\label{eq:const}
\end{eqnarray}
The above equations correspond to the asymptotic Bethe-ansatz equations
obtained by Kawakami\cite{kawa91}.  Our result thus confirms that the ABA
spectrum is exact.

Here the  resulting $\{p_\mu\}$ and $\{q_i\}$ must lie between $-\pi$
and $\pi$.  The set of $M+Q$ distinct quantum numbers $J_{\mu}$ is in
ascending order and governs the spin excitations.  We restrict them to
take values in the range $[-(N-M-1)/2,(N-M-1)/2]$ to guarantee that
they generate fully spin-polarized states.  There are $N-M$ values in
this range, of which $M+Q$ are occupied and $2S_z$ are empty.  A spin
configuration can be represented by a sequence of $N-M$ digits such as
$\{S\}=(0111001011)_s$, where 1 represents an occupied quantum number
and 0 an unoccupied quantum number.  These empty values are identified
as spinons\cite{hald91}; a sequence of $2j_r$ consecutive zeros
corresponds to a symmetric bound-state of $2j_r$ spinons, thereby
creating an excitation of spin $j_r$ with spin degeneracy $2j_r+1$. On
these physical grounds, we anticipate a spin degeneracy in the
thermodynamic limit given by
\begin{equation}
w_S= \prod_{j} (2j+1)^{n(j)}
\label{eq:degeneracy}
\end{equation}
where $n(j)$ is the number of sequences of zeros of length $2j$.  The
set of $Q$ distinct quantum numbers $I_i$, in ascending order and taking
values in the range $[-(M+Q)/2, (M+Q)/2]$, governs charge excitations.

To complete the study of the model and confirm our interpretation of
the quasiparticle degeneracies, we looked at exact diagonalization of
small systems ($N\le10$ with holes).  As an example, the low-lying states
of the $N=10$, $Q=2$ model is shown in Fig.~\ref{figone}.
We summarize the numerical result as follows:
\begin{enumerate}
\item The spectrum described in terms of the real pseudomomenta $\{p_\mu\}$
and $\{q_i\}$ span the $\it full$ set of energy levels of the system.
\item The real pseudomomentum states are all highest weight states when
$\{p_\mu\}\ne \pm\pi$.
\item The spin degeneracy rule is obeyed for all internal sequences of
zeros.
\end{enumerate}
Certain small corrections to the spin degeneracy rule apply when there are
zeros at either end of $\{S\}$ which we shall not enumerate here,
and which are not important in the thermodynamic limit\cite{unpub}.

\def\eps{\epsilon}

Finally, we may use the spectrum generated by the ``fully-polarized spinon
states'' and the supermultiplicity rule to obtain the free energy
of the model in the thermodynamic limit.
 Besides the ``particle-state''
solutions of equations (\ref{eq:eigen}) and (\ref{eq:const}),
we have to take into account the ``hole-state'' solutions\cite{yang71}.
At thermal equilibrium, the distribution functions
of these solutions are determined by minimizing the free energy
functional\cite{taka71,babu83}, $F = E - T S - \mu (N -Q)$,
with the constraint
that each ``fully-polarized spinon state'' described by quantum
numbers $\{ J_{\mu} \}$ and $\{ I_i \}$ is associated with
a spin degeneracy $w_S$ as given in (\ref{eq:degeneracy}),
where $\mu$ is the chemical potential.
Following the standard methods of Takahashi\cite{taka71}, minimizing
the free energy for a given quantum number distribution
gives the following free energy
\begin{equation}
F(T)/N=-\mu-{T\over2\pi}
\int_{-\pi}^{\pi} dp\,\ln \left[ 1 + e^{-\beta \eps_s (p) } \right],
\end{equation}
where $\epsilon_s$ is determined by the coupled equations
\begin{eqnarray}
2 \eps_s(p) &=& \eps_0(p) -2a - T \ln \left[ 1 + e^{-\beta\eps_c(p)}
 \right]\nonumber\\
\eps_c(q) &=&2a - T \ln \left[ 1 + e^{ -\beta\eps_s (q)}
\right].
\end{eqnarray}
Here $\eps_0 (p) ={1\over 2}t(p^2 -{\pi^2\over 3})+\mu$ and
$a={1 \over 6}t \pi^2 + {1 \over 2} \mu$. In the limit
of half filling, $\mu\rightarrow \infty$, $\eps_s\rightarrow t(p^2 - \pi^2)
/4$,
 and
the free energy reverts to the form obtained by Haldane for the corresponding
Heisenberg model\cite{hald91,tsvelik}.
For general $\mu$, elimination of $\eps_c(p)$ yields the result
\begin{equation}
\eps_s(p)=
\eps_o(p)-T
\ln \left[ {1 \over 2} +
\left({1 \over 4} + 2e^{\beta (\eps_0(p)+a)} \cosh (\beta a )\right)^{1/2}
\right],\quad
\end{equation}
We have verified
that  high temperature expansion of this free energy in powers of
$\beta$ correctly reproduces the first two non-trivial terms in the
high temperature perturbation theory.

{}From the free energy, it is not clear whether we may make
a unique identification of the statistics of the spin
and charge excitations. We note that the $S=1/2$
spinon excitations always combine into a state with a symmetric
spin wavefunction, thus $2S$ spinons form a state with total
spin $S$. In the limit of zero doping, the free energy is that of
spinless fermions\cite{hald91,tsvelik}. We can equally well
regard the spinon excitations  as $S=1/2$ fermions
in a state with a fully antisymmetric spatial wavefunction; or
alternatively, as hardcore $S=1/2$ bosons in  a
fully symmetric spatial state.


In summary, we have derived the spectrum of the 1D t-J model with
$1/r^2$ long-range exchange and hopping by the introduction
of zeros into Jastrow ground-state wavefunctions.
Our solution confirms  Kawakami's conjecture that the ABA
provides the exact spectrum, suggesting that
despite the long-range nature of the interactions,
two-body scattering dominates the long-wavelength physics.
By interpreting multiple occupancy of momentum states
in the spinon wavefunction as
symmetric bound-complexes of spinons, we have been able to determine
the degeneracies of the states needed to construct the free energy.
Further work is required to determine
the 	integrability conditions of this model.
There are also several possible generalizations: most notably,
SU(N) generalizations and the appealing
possibility of
Jastrow-integrable impurity models.

We would like to thank Z. Ha and F. D. M. Haldane for informative
discussions.
This work was supported in part by the U.S.~Department of Energy under
Grant No.~DE-FG05-85ER-40219, the National Science Foundation
under Grant NSF-DMR-13692, and by a Sloan Foundation Grant.

\figure{Low-lying energy levels of the 10 site 2 hole system from exact
diagonalization.  The numbers associated with each state list the
spin degeneracies starting with spin 0 on the left.
For example, the number ``331'' indicates that we have 3 states with
$S=0$,  3 states with $S=1$ and 1 state with $S=2$.
\label{figone}
}
\end{document}